\documentclass[aps,prl,twocolumn,superscriptaddress,showpacs,floatfix]{revtex4-1}

\usepackage{hyperref}
\usepackage{color}
\usepackage{graphicx}	
\graphicspath{%
  {./},%
}

\usepackage{amstext}
\usepackage{amsmath}
\usepackage{array}
\usepackage{enumitem} 
\usepackage{bm}
\usepackage{upgreek}

\usepackage{xspace}	

\newcommand{\vex}[1]{\vec{\mathbf{#1}}}
\newcommand{\vexs}[1]{\vec{\boldsymbol{#1}}}

\newcommand{\xperp}{\vec{\mathbf{x}}_\perp}

\begin{document}

\title{One fluid to rule them all: viscous hydrodynamic description of event-by-event central p+p, p+Pb and Pb+Pb collisions at $\sqrt{s}=5.02$ TeV}

\author{Ryan D.~Weller$^{1}$ and Paul Romatschke} 
\affiliation{Department of Physics, University of Colorado, Boulder, Colorado 80309, USA}
\affiliation{Center for Theory of Quantum Matter, University of Colorado, Boulder, Colorado 80309, USA}
\date{\today}

\begin{abstract}
  The matter created in central p+p, p+Pb and Pb+Pb collisions at $\sqrt{s}=5.02$ TeV is simulated event-by-event using the superSONIC model, which combines  pre-equilibrium flow, viscous hydrodynamic evolution and late-stage hadronic rescatterings.  Employing a generalization of the Monte Carlo Glauber model where each nucleon possesses three constituent quarks, superSONIC describes the experimentally measured elliptic and triangular flow at central rapidity in all systems using a single choice for the fluid parameters, such as shear and bulk viscosities. This suggests a common hydrodynamic origin of the experimentally observed flow patterns in all high energy nuclear collisions, including p+p.

\end{abstract}

\maketitle

\section{Introduction}

What are the properties of the matter created in ultrarelativistic ion collisions? Obtaining an answer to this question has been one of the key goals of the high energy nuclear physics community and the driving force behind the experimental heavy-ion program at both the Relativistic Heavy Ion Collider (RHIC) and the Large Hadron Collider (LHC). Much progress has been made, such as the realization that the matter created in heavy-ion collisions behaves more like a strongly interacting fluid, rather than a gas of weakly-interacting quarks and gluons \cite{Adcox:2004mh,Arsene:2004fa,Back:2004je,Adams:2005dq}. Some properties of this strongly interacting QCD fluid, such as the shear viscosity coefficient and the local speed of sound, have since been constrained \cite{Romatschke:2007mq,Song:2007ux,Dusling:2007gi,Bazavov:2009zn,Schenke:2010rr,Borsanyi:2013bia}. Others, such as the minimum possible size for a QCD fluid droplet, remain yet to be unambiguously determined. Before the present decade, the mainstream expectation was that a strongly interacting QCD fluid could only be formed in ``large'' systems, such as those created in heavy-ion collisions. ``Small'' systems, such as those formed in proton+nucleus or proton+proton collisions, were not expected to flow. It thus came as a surprise to many when experimental data from proton+nucleus and proton+proton collisions both at RHIC and the LHC unambiguously demonstrated the existence of flow in these small systems \cite{Abelev:2012ola,Aad:2012gla,Adare:2013piz,Adare:2015ctn,Aad:2015gqa}. 

The focus of the theory community has since shifted towards
understanding the origin of experimentally observed flow signals in small systems. 
At present, there are two main schools of thought. One maintains that while experimental evidence leaves no doubt that there are flow-like signals in small systems, these signals are unrelated to those observed in heavy-ion collisions and are caused by either initial-state correlations \cite{Dumitru:2014yza,Schenke:2015aqa,Altinoluk:2015uaa,Esposito:2015yva,Lappi:2015vta,Schenke:2016lrs,Hagiwara:2017ofm}, or non-hydrodynamic evolution, or non-standard final-state interactions \cite{Zhou:2015iba,Koop:2015wea,Bozek:2015swa,He:2015hfa,Romatschke:2015dha} or a combination of these. The other school of thought, on which the present work will be based, adheres to Heraclit's principle of ``Panta Rhei'' (``Everything Flows''). According to Panta Rhei, there is no fundamental difference between the experimental flow signals in small and large systems, and both can be quantitatively explained using the laws of hydrodynamics. (See Refs.~\cite{Romatschke:2016hle,Spalinski:2016fnj} for a discussion of why non-equilibrium hydrodynamics may be applicable to QCD fluid droplets as small as 0.15 fm).
Previous work on this subject includes the prediction of flow signals in p+p \cite{Ortona:2009yc,Prasad:2009bx,Werner:2010ss, Bozek:2011if}, p+Pb \cite{Bozek:2011if}, $^3$He+Au  \cite{Nagle:2013lja,Schenke:2014zha}, p+Au and d+Au collisions \cite{Romatschke:2015gxa} as well as the hydrodynamic description of experimental data in small systems (see e.g. Refs.~\cite{Kozlov:2014fqa,Habich:2015rtj} for recent examples).

One of the main criticisms of the Panta Rhei approach to relativistic ion collisions has been that a hydrodynamic description of experimental data of one or two individual collision systems could be a coincidence, and that a simultaneous description of small and large systems is required. Indeed, with the notable exception of Ref.~\cite{Werner:2014xoa}, previous hydrodynamic studies have focused on describing either proton+proton, proton+nucleus, or nucleus+nucleus collisions individually, rather than all of those systems together, which provides the motivation for the present study.

\section{Model}

The present study will be based on the super-hybrid-model superSONIC \cite{Habich:2014jna,Romatschke:2015gxa}, which combines pre-equilibrium dynamics (based on AdS/CFT \cite{vanderSchee:2013pia}) with viscous fluid dynamic evolution (\cite{Romatschke:2007mq,Luzum:2008cw}) and late-stage hadronic rescatterings (using the hadronic cascade code B3D \cite{Novak:2013bqa}). The superSONIC model has been used in the past to successfully predict experimental flow results in p+Au, d+Au and $^3$He+Au collisions at $\sqrt{s}=0.2$ TeV \cite{Aidala:2016vgl}. The main addition to the superSONIC model implemented here are initial conditions which allow for nucleon substructure. For the present work, a variant of the constituent quark model for the nucleon substructure will be used. This model is rather simplistic, and much more sophisticated models for nucleon substructure have previously been discussed in the literature, cf. Refs.~\cite{Avsar:2010rf,Yan:2013laa,Schenke:2014zha,McGlinchey:2016ssj,Bozek:2016kpf}. Given that hydrodynamics efficiently dampens small-scale fluctuations \cite{Noronha-Hostler:2015coa}, differences between models are expected to quickly dissipate, and it is likely that other nucleon models will give almost identical results to the constituent quark model employed in this work, as long as some basic granularity of the nucleon is maintained.
Hence, it is unlikely that the present work can be used to constrain nucleon structure models, with the possible exception of broad features such as the event-by-event ``ellipticity'' of the nucleon.

\subsection{Initial Conditions}

\begin{figure*}[t]
  \includegraphics[width=0.9\linewidth]{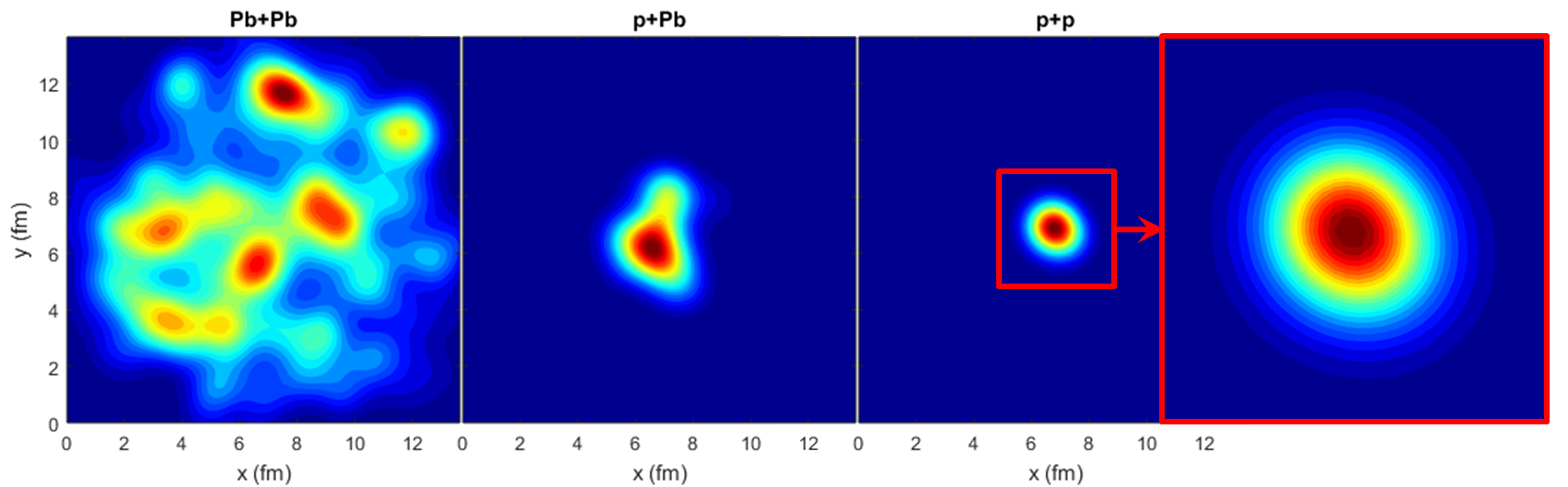}
\caption{Snapshots of typical energy density profiles in the transverse plane for Pb+Pb (left panel), p+Pb (center panel) and p+p collisions (right panel, including zoom-in to enlarge system)  at $\sqrt{s}=5.02$ TeV. The actual box sizes used in simulations were adapted to individual systems. Note that for a typical p+p collision, initial conditions from the OSU model are very close to (but nevertheless slightly different from) those obtained from spherical nucleons, cf. Ref~\cite{Habich:2015rtj}.  
\label{fig:one}}
\end{figure*}

The superSONIC model requires as input initial conditions for the energy density in the transverse plane; this density should correspond to the energy deposition following the collision of two relativistic nuclei. Each nucleus consists of individual nucleons whose positions and binary collisions are treated within a Monte Carlo Glauber framework. In addition, nucleon substructure is implemented through a type of quark model of the nucleon by modifying the work of Ref.~\cite{Welsh:2016siu}. This model shall be referred to as the OSU (Ohio State) model, and is used to generate event-by-event initial conditions as follows:

\begin{enumerate}[label=(\roman*)]
\item
The transverse positions of a beam and target nucleus are produced, separated by a random impact parameter of magnitude $b$. The impact parameter is obtained by sampling $b^2$ from a uniform distribution on the interval $[0,b_{\textnormal{cutoff}}^2]$, where $b_{\textnormal{cutoff}}$ is chosen to be large enough that the probability of a collision occurring is negligible when $b>b_{\textnormal{cutoff}}$. For our simulations, we set $b_{\textnormal{cutoff}}$ to 8, 20, and 40 fm for p+p, p+Pb, and Pb+Pb collisions, respectively.

\item If either the beam or the target nucleus is simply a proton, then a single nucleon is placed at the center of that nucleus. Otherwise, if a given nucleus is a lead nucleus, then the 3D positions of the nucleons comprising the nucleus are sampled from a Woods-Saxon distribution, using a radius of 6.62 fm and a skin depth of 0.546 fm \cite{DeVries1987495}. The transverse positions of the nucleons are then found by projecting their 3D positions onto the transverse plane.

\item Each nucleon in each nucleus is divided into three valence quarks. For a given nucleon $i$, the transverse positions $\vec{\mathbf{x}}_j^{(i)}$ ($j=1,2,3$) of the nucleon's valence quarks are generated in the following manner. First, two vectors $\vexs{\upchi}_1^{(i)}$ and $\vexs{\upchi}_2^{(i)}$ are sampled from Gaussian distributions, for which the respective probability density functions are $P_{1}(\vexs{\upchi}_1)=\frac{3}{2 \pi} e^{-3\vexs{\upchi}_1^2/2}$, $P_{2}(\vexs{\upchi}_2)=\frac{1}{2 \pi} e^{-\vexs{\upchi}_2^2/2}$.
Then, letting $\vex{x}^{(i)}$ denote the transverse position of the nucleon, the quark positions are set to
\begin{equation} \label{quark positions}
\begin{split}
\vec{\mathbf{x}}_1^{(i)}=\vec{\mathbf{x}}^{(i)}+\sqrt{3(B-\sigma_g^2)} (\vexs{\upchi}_1^{(i)} +\vexs{\upchi}_2^{(i)})/2, \\
\vec{\mathbf{x}}_2^{(i)}=\vec{\mathbf{x}}^{(i)}+\sqrt{3(B-\sigma_g^2)} (\vexs{\upchi}_1^{(i)} -\vexs{\upchi}_2^{(i)})/2, \\ 
\vec{\mathbf{x}}_3^{(i)}=\vec{\mathbf{x}}^{(i)}-\sqrt{3(B-\sigma_g^2)} \vexs{\upchi}_1^{(i)}.
\end{split}
\end{equation}
Here, the parameter $B$ controls the size of the nucleons, while the parameter $\sigma_g$ is a measure of the width of the Gaussian-shaped gluon cloud surrounding each valence quark \cite{Welsh:2016siu}. We set $B=(0.52\textnormal{ fm})^2$ and $\sigma_g=0.46\textnormal{ fm}$.

\item For all possible pairs $(i,j)$ consisting of a beam nucleon $i$ and a target nucleon $j$, a nucleon-nucleon overlap $T_{NN}^{(i,j)}$ is computed as
\begin{equation}
T_{NN}^{(i,j)}=\sum_{k,l=1}^{3} \frac{1}{4 \pi \sigma_g^2} e^{-| \vec{\mathbf{x}}_k^{(i)}-\vec{\mathbf{x}}^{(j)}_l|^2/(4 \sigma_g^2)},
\end{equation}
and the probability that a binary collision occurs between the two nucleons is determined as
\begin{equation}
P_\textnormal{collision}^{(i,j)}=1-e^{-\sigma_{gg} T_{NN}^{(i,j)}},
\end{equation}
where $\sigma_{gg}$ is a parameter that is fixed by requiring that across a large number of generated proton+proton events, the expected probability that a collision occurs satisfies
$\langle P_\textnormal{collision} \rangle \pi b_{\textnormal{cutoff}}^2=\sigma_{NN}^{\textnormal{inel}}=70\textnormal{ mb}$. Now, based on the probability $P_\textnormal{collision}^{(i,j)}$, it is randomly determined whether a binary collision will occur between nucleons $i$ and $j$ . If a binary collision occurs, nucleons $i$ and $j$ are both marked as ``wounded'', and become participants in the collision. Only nucleons, and not individual constituent quarks are marked as ``wounded'', which differs from other implementations of constituent quark Monte Carlo Glauber models in the literature \cite{Mitchell:2016jio}.

\item  Every wounded nucleon deposits entropy into the collision in the form of the gluon clouds surrounding its three valence quarks. (However, in the case that no nucleons are wounded, the procedure returns to step (i)). The entropy density $s(\vec{\mathbf{x}}_\perp)$ at a point $\vec{\mathbf{x}}_\perp=(x,y)$ in the transverse plane of the collision is then given by
\begin{equation} \label{s dens}
s(\xperp)=\kappa\sum_{n=1}^{N_w} \sum_{j=1}^{3} \gamma^{(n)}_j  \frac{1}{2 \pi \sigma_g^2} e^{-| \xperp-\vec{\mathbf{x}}^{(n)}_j|^2/(2 \sigma_g^2)},
\end{equation}
where $n$ is an index over all the wounded nucleons and $N_w$ is the total number of wounded nucleons. $\kappa$ is a parameter whose value is chosen to reproduce final-state charged hadron multiplicities, and $\gamma^{(n)}_j$ is proportional to the amount of entropy deposited near midrapidity by the $j^\textnormal{th}$ quark of the $n^\textnormal{th}$ wounded nucleon.  $\gamma^{(n)}_j$ is allowed to fluctuate from quark to quark with a probability density function 
\begin{equation} \label{gamma} 
P_\Gamma(\gamma)=\frac{\gamma^{1/(3\theta)-1} e^{-\gamma/\theta}}{\theta^{k}\Gamma(1/(3\theta))},
\end{equation}
where $\theta=0.75$ \cite{Welsh:2016siu}.

\item In the last step, the continuum entropy density profile (\ref{s dens}) is  converted to an energy density $\epsilon({\bf x}_\perp)$ using a lattice QCD equation of state \cite{Borsanyi:2013bia} and discretized on a square lattice adapted to the size of the collision system under consideration.

\end{enumerate}

Using the procedure described above, many initial energy-density profiles have been generated for p+p, p+Pb, and Pb+Pb collisions. For each initial profile, there is an associated total entropy per unit rapidity $dS/dy\propto\kappa\sum_{n=1}^{N_w}\sum_{j=1}^{3}\gamma^{(i)}_j$. Since $dS/dy$ increases with the total multiplicity of charged hadrons produced in a collision \cite{Song:2008si}, all initial density profiles are ordered into centrality classes based on their values for $dS/dy$. A subset of 100 initial conditions are randomly selected from each centrality class for further processing with superSONIC. Examples for typical transverse energy density profiles for central p+p, p+Pb and Pb+Pb collisions are shown in Fig.~\ref{fig:one}.

\begin{figure*}[t]
  \includegraphics[width=\linewidth]{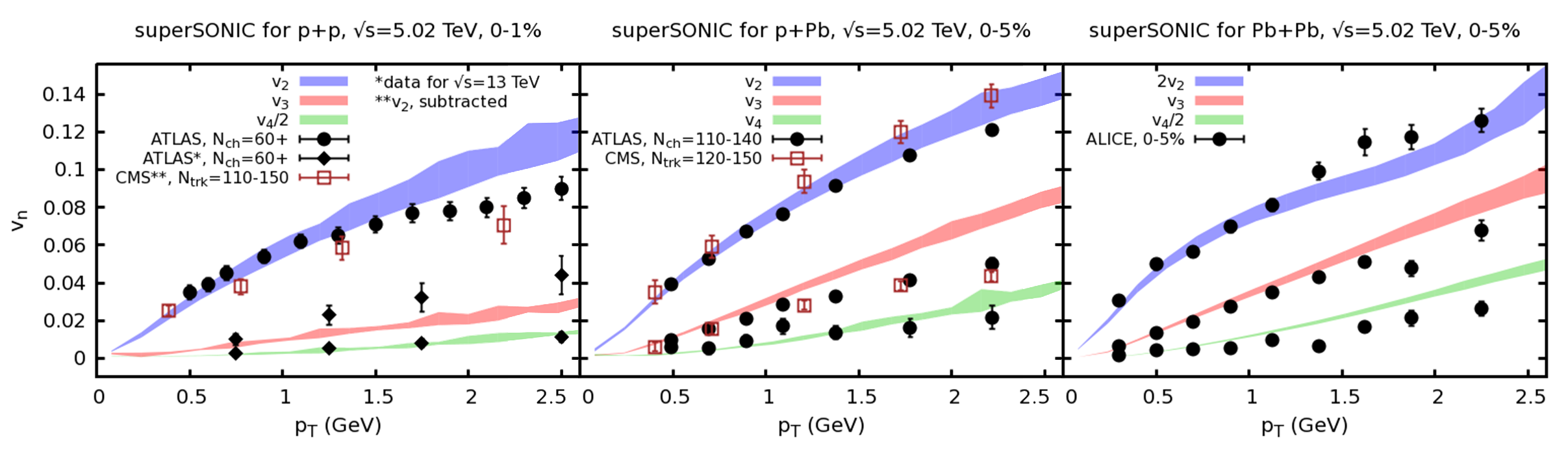}
\caption{Elliptic ($v_2$), triangular ($v_3$) and quadrupolar ($v_4$) flow coefficients from superSONIC simulations (bands) compared to experimental data from ATLAS, CMS and ALICE (symbols) for p+p (left panel), p+Pb (center panel) and Pb+Pb (right panel) collisions at $\sqrt{s}=5.02$ TeV \cite{Khachatryan:2016txc,Aad:2014lta,Adam:2016izf,Aaboud:2016yar, Chatrchyan:2013nka}. Simulation parameters used were $\frac{\eta}{s}=0.08$ and $\frac{\zeta}{s}=0.01$ for all systems. Note that ATLAS results for $v_3,v_4$ are only available for $\sqrt{s}=13$ TeV, while all simulation results are for $\sqrt{s}=5.02$ TeV. 
\label{fig:two}}
\end{figure*}

\subsection{superSONIC}

The superSONIC model converts initial energy density profiles into spectra of identified particles that can directly be compared to experimental data (see Ref.~\cite{Romatschke:2015gxa} for a more detailed description of the model). In brief, for each initial energy-density profile $\epsilon({\bf x}_\perp)$, a pre-equilibrium flow profile at proper time $\tau=\sqrt{t^2-z^2}$ is generated using $\vec{v}(\tau,{\bf x})=-\frac{\tau}{3.0}\vec \nabla \ln \epsilon({\bf x}_\perp)$ \cite{Vredevoogd:2008id}, consistent with gauge/gravity simulations of strongly coupled matter \cite{vanderSchee:2013pia}, while the value of the shear and bulk stress tensors will be set to zero. Using these initial conditions, 2+1 dimensional hydrodynamic simulations at mid-rapidity are then started at time $\tau=\tau_0=0.25$ fm using a lattice QCD equation of state \cite{Borsanyi:2013bia} and  shear and bulk viscosity values of $\frac{\eta}{s}=0.08$ and $\frac{\zeta}{s}=0.01$, respectively. Bulk viscous effects on particle spectra are at present poorly understood \cite{Monnai:2009ad} so only effects of bulk viscosity on the hydrodynamic evolution is included, and for simplicity bulk and shear relaxation times are identical, $\tau_\Pi=\tau_\pi$ \cite{Kanitscheider:2009as}. The corresponding shear viscous relaxation time is varied between $\tau_\pi=4 \frac{\eta}{s T}$ and $\tau_\pi=6 \frac{\eta}{s T}$ in order to quantify the sensitivity of results to non-hydrodynamic modes \cite{Romatschke:2016hle}, where $T$ denotes the local effective temperature of the system. Large variations of observables with $\tau_\pi$ are indicative of a breakdown of hydrodynamics, while small variations suggest that hydrodynamics still applies as an effective bulk description.
Simulations were performed on lattices with $100\times 100$ grid points, with lattice spacings adapted to the individual size of the collision system (cf. Fig.~\ref{fig:one}). In addition, test simulations with $200\times 200$ gridpoints were used to ensure that finite volume and finite resolution artifacts are under control. Once the local temperature reaches $T=0.17$ GeV in a given fluid cell, hydrodynamic variables and location of the cell are stored for further processing using the low-temperature hadronic cascade evolution with B3D \cite{Novak:2013bqa}. B3D simulates the s-wave scatterings with a constant cross section of 10 mb and interactions through hadron resonances in the particle data book with masses up to $2.2$ GeV. After resonances have stopped interacting, the final charged particle multiplicity as well as hadron spectra are obtained, and can be directly compared to experimental measurements at mid-rapidity.
The source code to superSONIC is publicly available \cite{codedown}.

\section{Results}

Using superSONIC with OSU initial conditions for the nucleon, central p+p, p+Pb and Pb+Pb collisions at $\sqrt{s}=5.02$ TeV have been simulated using one single fluid framework with fixed values of shear and bulk viscosity coefficients for all systems. The results for the differential elliptic, triangular and quadrupolar flow at midrapidity from superSONIC are shown in Fig.~\ref{fig:two} together with experimental results from the ALICE, CMS and ATLAS experiments \cite{Khachatryan:2016txc,Aad:2014lta,Adam:2016izf,Aaboud:2016yar, Chatrchyan:2013nka}. The size of the bands shown for superSONIC calculations includes statistical errors for the simulations as well as systematic uncertainties obtained from changing the second-order transport parameter $\tau_\pi$. The size of the uncertainty bands suggests that simulation results for all systems shown are not strongly sensitive to the presence of other, non-hydrodynamic modes, and thus a hydrodynamic effective description seems applicable.

Overall, Fig.~\ref{fig:two} implies good agreement between the superSONIC model and experiment at low momenta for all collision systems when taking into account the systematic and statistical uncertainties in both the theory and experimental results. It should be pointed out that no fine-tuning of superSONIC parameters has been attempted, so no precision fit of the experimental data can be expected. Furthermore, note that in the case of p+p collisions, ATLAS data for $v_3,v_4$ is only available for $\sqrt{s}=13$ TeV, more than twice the simulated collision energy of $\sqrt{s}=5.02$ TeV.

\begin{figure*}[t]
  \includegraphics[width=.32\linewidth]{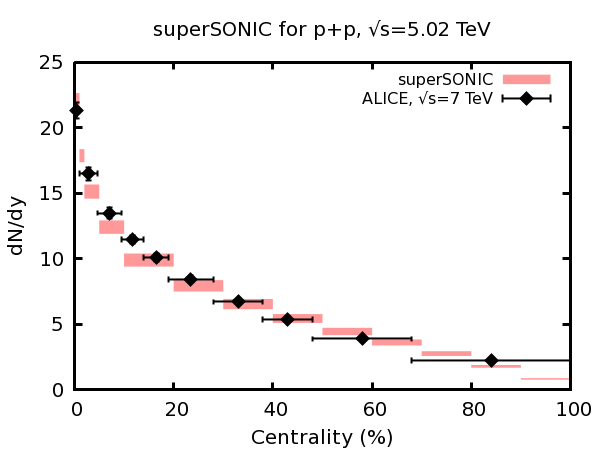}
  \includegraphics[width=.32\linewidth]{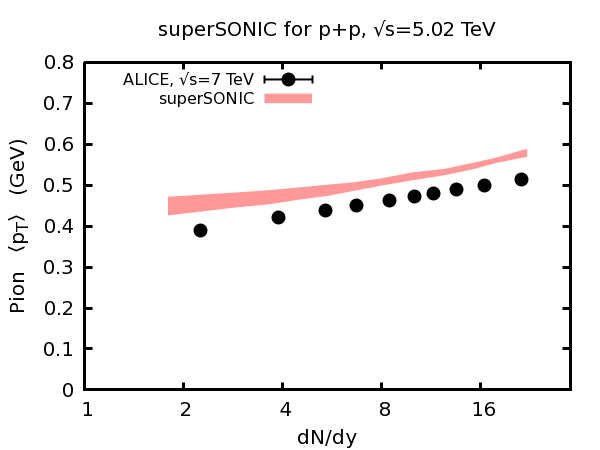}
  \includegraphics[width=.32\linewidth]{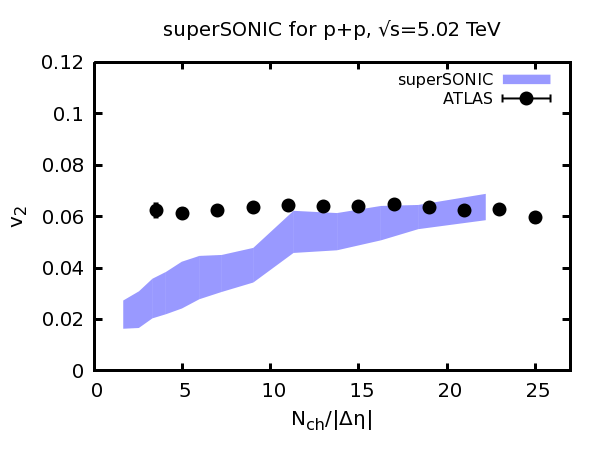}
  \caption{Multiplicity, pion mean transverse momentum and integrated elliptic flow coefficient for p+p collisions at $\sqrt{s}=5.02$ TeV from superSONIC (bands) compared to experimental data from ALICE at $\sqrt{s}=7$ TeV and ATLAS at $\sqrt{s}=5.02$ TeV (symbols) \cite{Adam:2016emw,LivioTalk}. Simulation parameters used were $\frac{\eta}{s}=0.08$ and $\frac{\zeta}{s}=0.01$ for superSONIC and multiplicities were converted to charged particles per unit pseudorapidity as $\frac{dN}{dy}=1.1 \frac{N_{\rm ch}}{|\Delta \eta|}$.
\label{fig:three}}
\end{figure*}

The case of p+p collision at $\sqrt{s}=5.02$ TeV has moreover been studied as a function of multiplicity, and results for the multiplicity, mean pion transverse momentum, and integrated elliptic flow are shown in Fig.~\ref{fig:three} together with experimental data. This figure suggests that the multiplicity distribution is well represented in the superSONIC model, while the pion mean transverse momentum only qualitatively matches experimental results: the simulated $\langle p_T\rangle$ values exceed the results measured by ALICE (at $\sqrt{s}=7$ TeV) at all multiplicities. This finding is not surprising given that present simulations did not include bulk viscous corrections to the pion spectra, which can be expected to considerably affect $\langle p_T \rangle$ results, cf. Refs.~\cite{Habich:2015rtj,Monnai:2009ad,Ryu:2015vwa}. Given the extreme sensitivity of $\langle p_T\rangle$ on bulk viscosity for proton+proton collisions \cite{Habich:2015rtj}, it is quite possible that including bulk corrections to spectra and/or fine tuning can lead to quantitative agreement of simulation and experiment for $\langle p_T\rangle$ in p+p collisions, while not significantly altering results for p+Pb and Pb+Pb collisions. Such fine-tuning is left for future work.

Also shown in Fig.~\ref{fig:three} is the integrated elliptic flow coefficient as a function of multiplicity, indicating that $v_2$ saturates at high multiplicities similar to what is observed experimentally. At low multiplicities, experimental procedures employed by different experiments lead to different results. So while the method employed by the ATLAS experiment suggests a near constant behavior of $v_2$ as a function of multiplicity, the method employed by CMS (not shown in Fig.~\ref{fig:three}) by construction implies that integrated $v_2$ decreases as multiplicity is lowered. Nevertheless, reproducing the apparent saturation of integrated $v_2$ at around 6 percent for high multiplicities (for which both ATLAS and CMS experiments agree on) is non-trivial for any model as this trend depends on the choice of shear viscosity and nucleon initial state parameters.

For p+Pb collisions and Pb+Pb collisions at $\sqrt{s}=5.02$ TeV, the model results for $\frac{dN}{dy}$ for the 0-5\% highest multiplicity events  are within five percent of the experimental values at midrapidity \cite{Aad:2015zza,Adam:2015ptt} when converting superSONIC multiplicities to pseudorapidity distributions as $\frac{dN}{dy}=1.1 \frac{dN}{d\eta}$.

\section{Conclusions}

Relativistic p+p, p+Pb and Pb+Pb collisions at $\sqrt{s}=5.02$ TeV and small impact parameter have been simulated event-by-event using the super-hybrid-model superSONIC. Using initial conditions that allow for nucleon substructure in the form of three valence quarks, it was found that flow in all collision systems can be described simultaneously with a single set of fluid parameters. This finding suggests that the experimentally observed flow signals in proton+proton, proton+nucleus and nucleus+nucleus collisions are of common, and hydrodynamic, origin. However, more work will be needed to corroborate this conclusion.

\begin{acknowledgments}
\section{Acknowledgements}
 
This work was supported in part by the Department of Energy, DOE award No DE-SC0008132 and by the UROP program at the University of Colorado Boulder. We would like to thank M.~Floris, M.~Habich, J.~Nagle and D.~Perepelitsa for fruitful discussions.

\end{acknowledgments}

\bibliographystyle{apsrev} \bibliography{pp-hydro}

\end{document}